\pdfoutput=1
\documentclass[11pt]{article}

\usepackage{emnlp2021}
\usepackage[frozencache,cachedir=.]{minted}

\usepackage{times}
\usepackage{latexsym}

\usepackage{dirtytalk}
\usepackage{amsfonts,amssymb}
\usepackage{booktabs}
\usepackage{newtxtext,newtxmath}
\usepackage{array, caption, threeparttable}
\usepackage{multirow}
\usepackage{graphicx}
\usepackage{float}

\usepackage[T1]{fontenc}
\usepackage[utf8]{inputenc}
\usepackage{microtype}

\title{HAConvGNN: Hierarchical Attention Based Convolutional Graph Neural Network for Code Documentation Generation in Jupyter Notebooks }

\author{
Xuye Liu $^{\ddagger}$  \\ University of Waterloo
\And Dakuo Wang $^{\ddagger}$  \\ IBM Research \And 
 April Yi Wang \\ University of Michigan
\AND 
 Yufang Hou \\ IBM Research Europe
 \And Lingfei Wu \thanks{$^{\ddagger}$ Equal contributions from the first authors: \texttt{x827liu@uwaterloo.ca, dakuo.wang@ibm.com}. Part of work was done when Xuye, April, and Lingfei were at IBM.} \\ JD.COM Silicon Valley Research Center}

\begin{document}
\maketitle
\begin{abstract}
Jupyter notebook allows data scientists to write machine learning code together with its documentation in cells. 
In this paper, we propose a new task of code documentation generation (CDG) for computational notebooks. In contrast to the previous CDG tasks which focus on generating documentation for single code snippets, 
%The automated code documentation generation (\texttt{CDG}) task in notebooks is relevant to code summarization in software engineering, but a novel task, 
%as 
in a computational notebook, one documentation in a markdown cell often corresponds to multiple code cells, 
and these code cells have an inherent structure. We proposed a new model (\emph{HAConvGNN}) that uses a hierarchical attention mechanism to consider the relevant code cells and the relevant code tokens information when generating the documentation. Tested on a new corpus constructed from well-documented Kaggle notebooks, we show that our model outperforms other baseline models.

% Our model (HAConvGNN) solves this new \texttt{CDG} task by encoding these multiple code cells each as an AST graph structure, and together as a higher-level graph, for which we propose a hierarchical attention mechanism to augment a Seq2Seq network. We also construct a dataset from well-documented Kaggle notebooks, and show that our model outperforms baseline models. 
\end{abstract}
\section{Introduction}
In recent years, computational notebooks such as Jupyter have become popular programming platforms for data scientists and machine learning researchers to document ideas, write code, and visualize results, all in a single document \cite{wang2021makes}.
% Data scientists and machine learning researchers commonly use computational notebooks to explore and experiment their model building or data preprocessing ideas, as the notebook can support them to combine code, result, and documentation together~\cite{ kery_exploring_2017}.
Documentation in a notebook provides a rich medium for users to record not only what the code does, but also why they code it. This richness of content is one distinctive nature of code documentation in a notebook versus in traditional software source code.

Code documentation is found critical for data scientists to share or reuse code~\cite{zhang2020data,chattopadhyay2020s}.  %but 
However, research %\cite{rule2018exploration}
%shows
has shown
that many data scientists still neglect to write appropriate documentation for their code in  notebooks, as they feel writing documentation will slow down their coding process. 
\newcite{rule2018exploration} report that among one million computational notebooks on Github,  25\% of them have no comment. 

\begin{table}[ht]
\centering
\small
\begin{tabular}{p{7cm}}
\\ \hline
\textbf{Documentation}
\\ \hline
\begin{tabular}{l l}
{\color{red} \textbf{ground truth}} & Implementing Neural Network \\
{\color{blue} \textbf{our Model}} & Implementing Neural Network \\
{\color{magenta} \textbf{code2seq}} & The following function of the model \\
{\color{olive} \textbf{graph2seq}} & After perturbations \\
{\color{cyan} \textbf{T5-small}} & Model \\
\end{tabular} 
\\ \hline
\textbf{Code Cells}
\\ \hline
\vspace{-12pt}
\begin{minted}[%
 breaklines,
 fontsize=\scriptsize
 ]{python}
import keras
from keras.utils import plot_model
from keras.models import Model,Sequential,load_model
...
\end{minted}
\\ \hline
\vspace{-12pt}
\begin{minted}[%
 breaklines,
 fontsize=\scriptsize
 ]{python}
def nn_model(X,y,optimizer,kernels):
    input_shape = X.shape[1]
       
    if(len(np.unique(y)) == 2):
        op_neurons = 1
        op_activation = 'sigmoid'
        loss = 'binary_crossentropy'
    else:
        op_neurons = len(np.unique(y))
        op_activation = 'softmax'
        loss = 'categorical_crossentropy'
    
    classifier = Sequential()
    ...
    
    classifier.summary()
    return classifier
\end{minted} 
\\ \hline
\vspace{-12pt}
\begin{minted}[%
 breaklines,
 fontsize=\scriptsize
 ]{python}
model = nn_model(X_train,y_train,'adam','he_uniform')
history = model.fit(X_train, y_train, batch_size = 64, 
                    epochs = 1000, validation_data=(X_test, y_test))
\end{minted} 
\\ \hline
\vspace{-12pt}
\begin{minted}[%
 breaklines,
 fontsize=\scriptsize
 ]{python}
pd.DataFrame(abs(train.corr()['Survived']).sort_values(
ascending = False))
\end{minted} 
\\ \hline
\end{tabular}
\caption{An example of multiple code cells after one documentation block}
\label{tab:example}
\end{table}

As a first step towards building an automated documentation generation system for notebooks, in this paper we focus on the code documentation generation (CDG)  task for Jupyter notebooks. Since there is no publicly available CDG dataset for notebooks, we construct a new dataset (\textbf{notebookCDG}) which contains around 28k processed code-documentation pairs extracted from 2,476 highly-ranked notebooks from Kaggle competitions (details in Section \ref{sec:dataset})

% Automated \texttt{CDG} technique may help. 
A few previous literature have explored techniques to generate documentation for software code snippet one at a time \cite{LeClair2019GNN, haque2020improved, haque2021action, xu2019graphseq}. However, in computational notebooks, 
%However, existing code summarization techniques have two limitations when applied to the context of computational notebooks:
%First, prior work (e.g.,~\cite{LeClair2019GNN, xu2019graphseq}) only consider one summary for one single code snippet.
%In computational notebooks, 
one documentation (in a \emph{markdown cell}) %may 
can cover %summarize or to rationalize 
more-than-one code cells after it. For instance, the ground truth text in Table \ref{tab:example} is a single documentation covering four code cells. 
%Second,
% Furthermore, the documentation in notebooks often has a linear?? hierarchical structure.
% For example,  data preparation related code cells and documentation cells are often ahead of the model training cells. 
Existing work on CDG \cite{kery_exploring_2017, Iyer, Hu, alon2018codeseq, LeClair2019GNN} does not consider such structure information since they only focus on documentation generation for single code snippet (i.e., one function, or one expression).

%Existing literature~\cite{LeClair2019GNN, xu2019graphseq} does not consider  the hierarchical structure of documentation or for more than one code snippets. 
% Therefore, we can consider the code documentation task in computational notebooks as a multi-document generation task. In our task, we have a two-level design in our model. At the lower level, we analyze every single code snippet in multiple code cells. At the higher level, we design a hierarchical structure for multiple code cells. We firstly apply a attention mechanism to multiple code cells. Then we apply a new attention mechanism to find important code tokens in each code cell.

To account for the above mentioned properties of documentation in computational notebooks, in this paper, we propose a graph-augmented encoder-decoder model to generate documentation for notebooks (Section \ref{sec:approach}).
%for the \texttt{CDG} task in %the computational notebook context. 
In particular, our model consists of three parts: a code sequence encoder, an auxiliary documentation text encoder based on the already predicted documentation tokens, and a Hierarchical Attention-based Convolutional Graph Neural Network (HAConvGNN) component. 

The first two sequence encoders encode the semantic information in code and documentation text, respectively. The graph encoder encodes the contextual abstract syntactic trees (i.e., AST extracted from the code sequence). 
%in order to better capture all structural information from the inputs. 
In order to capture the relations between code sequences and the corresponding text documentations, we 
%then
further employ a hierarchical attention mechanism consisting of a low-level attention module and a high-level attention module. The former attends to the token in a code sequence and the latter attends to the corresponding code cells in the AST tree. 
%in order to learn a stronger relationship between codes and the predicted text sequence.

%We compare our model and a few other baselines 
%A comparison between our model and the other two baseline models 
Experiments show that our model achieves  better performance on the \emph{notebookCDG} dataset compared to baseline models on ROUGE scores, and in a muti-dimensional human evaluation study. 
% In the human rating evaluation study, we found that our model outperforms other baselines in different dimensions. 

Base on this result, we integrated our approach into a user-facing downstream application~\cite{wanggraph} to further explore the Human-AI collaboration opportunity in the code documentation scenario. 
In the follow-up user study (reported seperately~\cite{wang2021themisto}), users found that the automatically generated documentation reminded them to document code they would have ignored, and improved their satisfaction with their computational notebooks.  %, improving  ROUGE-LCS F1 score (18.54) by over 81\% compared to the code2seq model (10.24) and over 62\% compared to the graph2seq model (11.18).

In summary, the main contributions of our work are: (1) a large-scale high quality dataset for the CDG task in the computational notebook context; (2) a graph-based neural network architecture with hierarchical attention for the notebook CDG task which considers the structure information between multiple code cells and the relations between code tokens and text tokens; and (3) human evaluations to validate our model for real world application. The experiment code and data are shared\footnote{\url{https://github.com/dakuo/HAConvGNN}}.
% at \url{https://github.com/Anonymized_for_review}.
\section{Related Work}

\begin{table*}[t]
\centering
% \begin{minipage}{0.48\linewidth}
\centering
%\small
\begin{tabular}{l| l| l|l| l}
\toprule
&\textbf{Overall}&\textbf{Train} &\textbf{Dev} &\textbf{Test}  \\
\midrule
Notebooks number & 2,476&2,426&1,390&1,394\\
Code-documentation pairs &28,625&22,851&2,856&2,856\\
Code vocabulary size  &20,522&&\\
Code AST vocabulary size  &67,211&&\\
Documentation vocabulary size  &13,053&&&\\ \hline
Avg. \# token in documentation &9.15&9.13&9.37&9.18\\
Max. \# token in documentation &202&202&130&104\\
Std. \# token in documentation &8.40&8.44&8.27&8.25\\ \hline
Avg. \# token in code cell(s) &65.38&65.50&65.41&64.39\\
Max. \# token in code cell(s) &400&400&400&395\\
Std. \# token in code cell(s) &68.93&69.16&68.23&67.71\\\hline
Avg. \# token in code AST &181.08&181.47&180.77&178.24\\
Max. \# token in code AST &1732&1548&1732&1167\\
Std. \# token in code AST &192.19&193.00&190.43&187.40\\
\bottomrule
\end{tabular}
\caption{\textit{\textbf{notebookCDG}} dataset statistics. The overall code-to-markdown ratio is 2.2195, which suggests one markdown corresponds to more than one code cells.}
\label{table:dataset}
% \end{minipage}

% \begin{minipage}{0.48\linewidth}
% \centering
% \small
% \begin{tabular}{c c}
% \toprule
% Unigram&\textbf{Frequency} \\
% \midrule
% We & 3.06\%\\
% The & 2.30\%\\
% Model & 1.36\%\\
% Feature & 1.31\%\\
% Data & 1.18\%\\
% I & 0.97\%\\
% There & 0.80\%\\
% Create & 0.79\%\\
% Random & 0.77\%\\
% Train & 0.73\%\\
% \bottomrule
% \end{tabular}
% \caption{ Frequent unigrams in the documentation in our
% dataset}
% \label{table:first_token}
% \end{minipage}
\end{table*}

% Documenting the story behind codes and results is critical for data scientists to collaborate with others, as well as with their future selves~\cite{kery_exploring_2017}. 
% Automated \texttt{CDG}
% has many potential applications such as for code search~\cite{howard2013automatically}. 
% Though the \texttt{CDG} task in computational notebook context is new, we can review related research work on general program code summarization task.
%\paragraph{Machine Learning for Code.}
In order to automate the machine learning and AI workflow, researchers have 
applied automation techniques on various  code-related tasks~\cite{wang2020autoai}, including code summarization \cite{Iyer, LeClair2019GNN,haque2020improved, haque2021action}, source code generation from natural language \cite{agashe-etal-2019-juice}, and source code transformation \cite{NEURIPS2020_ed23fbf1}.

In this work, we focus on the code documentation generation(CDG) task. Our work is closely related to code summarization. 
Most existing datasets for code summarization contain one summary per one code snippet. For instance, CodeSearchNet \cite{DBLP:journals/corr/abs-1909-09436} contains two million function-documentation pairs across six programming languages (e.g., java, php, python).
In contrast, our new dataset (\emph{notebookCDG}) is designed for computational notebooks. The difference from previous CDG datasets is that in our dataset, a documentation text can correspond to several code snippets.

Previous work on code summarization focuses on summary generation for a single standalone code snippet. 
~\newcite{Iyer} collected Stack Overflow question titles as code summaries and paired them with top-rated code snippets. They then used an attention seq2seq model to generate a summary for each code snippet.
Several studies explored the abstract syntactic tree (AST) information of source code to better capture the relation between different elements \cite{Hu,alon2018codeseq}. 
Recently,  \newcite{xu2019graphseq} and \newcite{chen2019reinforcement} have proposed a general graph to sequence model to learn node embeddings and then reassemble them into the graph embeddings. 

Unlike the aforementioned works that only focus on summary generation for a single standalone code snippet, 
in our new \texttt{CDG} task for computational notebooks, multiple adjacent code cells can correspond to one documentation and these code cells may have a hierarchical structure, and use a graph to represent it~\cite{kipf2016semi}. We thus 
propose Hierarchical Attention-based Convolutional Graph Neural Network (HAConvGNN)
to handle the hierarchical AST graph structure of multiple code cells.

\section{\textit{notebookCDG} Dataset}
\label{sec:dataset}

\texttt{CDG} for notebooks is a relatively new task. To our best knowledge, we could not find an appropriate dataset for this task. Thus, we decided to construct a new dataset and share it with the community.
Publicly shared notebooks on Github are often ill-documented~\cite{rule2018exploration}, thus are not suitable for constructing the training dataset for CDG task.
A recent work~\cite{wang2021makes} manually analyzed 80 publicly available notebooks on two Kaggle challenges (i.e. out of 12,000 notebooks submitted to Titanic and HousePrice). 
Kaggle allows community members to vote up and down on those notebooks, and ~\newcite{wang2021makes}'s findings show that the highly-voted notebooks are of good quality and quantity in code documentation.
% Also, they found that on average there are 55.3 code cells and 45.1 markdown cells in each notebook, which implies one markdown cell is corresponding to multiple code cells.
Inspired by their work, we decided to utilize the top-voted and well-documented Kaggle notebooks to construct the \textbf{notebookCDG} dataset\footnote{We share the notebookCDG dataset with processed 28k code-document pairs at \url{https://ibm.biz/Bdfpk6} 
% \url{https://ibm.biz/BdfpkU}
}.

We collected the top 10\% highly-voted notebooks from the top 20 popular competitions on Kaggle (e.g. Titanic). We checked the data policy of each of the 20 competitions, none of them has copyright issues.
We also contacted the Kaggle administrators to make sure our data collection complies with the platform's policy. In total, we collected 3,944 notebooks as raw data.

\subsection{Data Preprocessing}
We performed various preprocessing steps to prepare the dataset, following~\newcite{Leclairdata}. For example, we removed notebooks in non-English language. 
One major difference between our dataset and previous datasets is that in previous datasets, each documentation unit is corresponding to one code snippet, whereas in our dataset, one documentation unit may correspond to upto four code snippets (code cells). 
% that when generating AST nodes and AST edges, we select four code blocks between pre-markdown and post-markdown. We treat each code block separately and generate AST for each code block, which is used for Hierarchical Attentional GNN based model training. 
We first located the markdown cells that have code cells beneath them. 
According to~\newcite{wang2021makes}, there are nine categories of documentations in a notebook, some are related to code, some are not related to code. For those types closely related to code (\texttt{Process} and \texttt{Headline}), which take up 80\% of the cases, we can directly use the markdown cell as documentation. For some other types, such as the \texttt{Result} type, which interprets the rendered result table or plot thus are often long and irrelevant to the code, we used a list of keywords (e.g., shows) to filter out the key sentences from the markdown cell as the documentation.
Another special types of documentation are \texttt{Reason} and \texttt{Education}, which also uses long word sequence to explain why the author did something.
In these cases, based on our observation, we used the first sentence as the documentation, as the first sentence is often related to the code cells.
% We used the preprocessed markdown cells' content as our target documentation sequence. 

Our analysis shows that for one markdown cell, there could have maximum four code cells following it. 
We construct our dataset to have a structure with one documentation unit and four code sequence units, and fill with empty sequence if there is less than four code sequences.
% In contrast, if the number of the code snippets is greater than four, we will extract the first four code snippets.
% For our baseline comparisons, we use a tokenized version of the dataset.
As part of the data preparation, we also parse each of code sequence to an AST graph structure through a Python AST library\footnote{\url{https://docs.python.org/library/ast.html}}. While doing so, we removed all the non-Python notebook magic (e.g. \%matplotlib).
% The difference between our dataset and their dataset is that in their data one summary matches to one code snippet, in ours one summary matches to one or multiple code snippets.
% We will share our raw and preprocessed dataset upon the paper acceptance.
% when generating AST nodes and AST edges, we selected four code blocks between pre-markdown and post-markdown. 
% We treat each code block separately and generate AST for each code block, which is used for Hierarchial Attentional GNN based model training. 

% When analyzing the content in the markdown cell, based on the classification of markdown cell by ~\cite{wang2021makes}, we find that markdown cells classified as reason and education will explain the function or purpose of the following code blocks in their first sentence, and then will start to describe some more detailed content which cannot be generated. 
% Therefore, because sometimes the markdown content is too long, we have two approaches to further shorten the documentation text: First, we identify the content enclosed by a headline schema, and use it directly as the target documentation text; if there is no headline schema, we then extract the first sentence in the markdown cell as the target documentation sequence. 
% At the same time, we also need to remove the markdown cell that cannot be generated by a code sequence, such as result type markdown cell~\cite{wang2021makes}. In this way, we use the method of finding keywords which denoting result such as so, show, represent, etc. to skip this kind of markdown cell.

\subsection{Dataset Core Statistics}
After data preprocessing, the final dataset contains 2,476 notebooks out of the 3,944 notebooks from the raw data. 
It has 28,625 code--documentation pairs. 
The overall code-to-markdown ratio is 2.2195, which suggests one markdown corresponds to more than one code cells.
Then, the code-documentation pairs are randomly split into train, dev, and test subsets, following a 8:1:1 ratio (Table \ref{table:dataset}). 
% This approach implies that pairs from the same notebook could be randomly assigned into either train, or dev, or test subset.

% The dataset contains 28,625 code(s)-documentation pairs. The code(s) are from notebooks we collected from the Kaggle competition. The documentations are extracted from pre-Markdown cells in these notebooks, average 9.15 tokens in length, in which the most common first token can be seen in Table \ref{table:first_token}. 

Our \textit{notebookCDG} dataset has a vocabulary size of 13,053 for the documentation sequence, a vocabulary size of 20,522 for the code sequence, and 67,211 for the parsed code AST node.
On average, each pair of code-documentation has 65.38 code tokens, and 9.15 documentation tokens. When code is translated to AST structure, on average it has 181.08 tokens.

\section{Approach}
\label{sec:approach}

Our model is built upon the standard encoder-decoder structure. To handle multiple code cells in computational notebooks, we propose a hierarchical attention mechanism based on convolutional graph neural network (HAConvGNN) for capturing the  relevant code cells during the decoding stage. 

\begin{figure*}[t]
    \includegraphics[width=16cm]{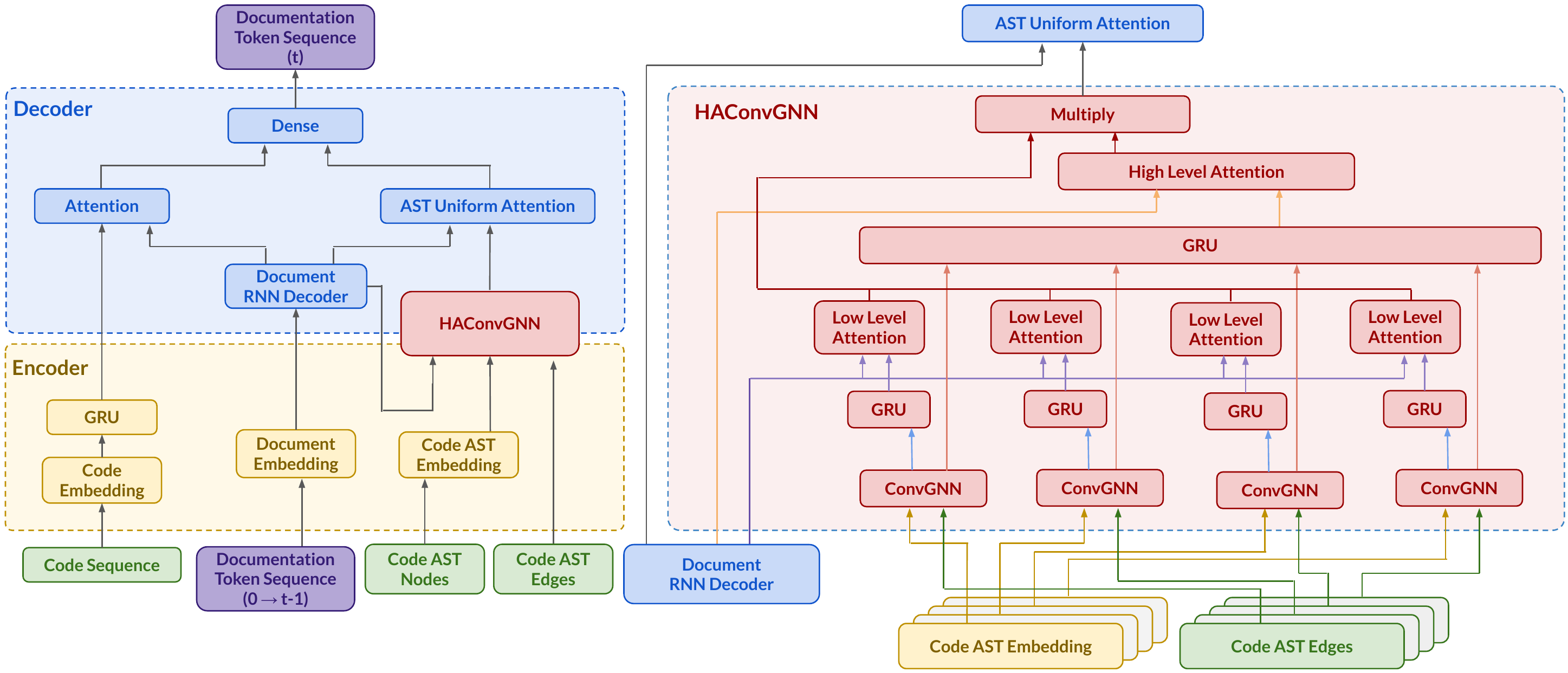}
    \caption{HAConvGNN model architecture}
    \label{fig:model}
\end{figure*}

The system architecture is illustrated in Figure \ref{fig:model}. Below, we describe each module in detail.
%In this section, we provide model details
% , inspired by~\cite{LeClair2019GNN}. Their work shows that Convolutional Graph Neural Networks are well suited in code summarization task. The method they use to combine the information from neighbor nodes is aggregation. We design a specific 
% for the Hierarchial Attentional Convolutional Graph Neural  Network (HAConvGNN).

\subsection{Model Input}
As mentioned in Section~\ref{sec:dataset}, we found that there are up to four adjacent code cells under a markdown cell, thus we constructed the \textit{notebookCDG} dataset to have one documentation mapping to four code cells, and used empty code cell as padding.
Therefore, when generating the abstract syntactic tree (AST) for a code cell, we can assemble up to four AST trees into a higher level graph structure. 

In summary, each training data point has four parts: the tokenized code sequence,  the tokenized documentation sequence, the nodes of the AST graph generated from the code sequence, and the edges (topology) of the AST graph generated from the code sequence. We denoted code sequence input as   $S =  \{{s_1},{s_2},...,{s_n}\}  \in \textbf{\textit{S}}$  where 
$s_i$ is sequence consisting of a sequence of code token embeddings $s_i$ = $\{w_1, w_2, ..., w_k\} \in \textbf{\textit{W}}$ in which \textbf{\textit{W}} is the token embedding space and $k$ is the length of $s_i$. Next we construct the AST graph input $A = (V, E)$ where V are the nodes containing the original code, %and E is the relationship structure from the parsed AST graph. 
${E}$ are the edges which denote whether two nodes are connected or not in the AST graph. 

\subsection{Embeddings}
We use three embedding layers to generate embeddings for the tokenized code sequence, the nodes in an AST graph, and the documentation decoder,  respectively. 

\subsection{Encoder}
We use one encoder to encode the source code sequence, and additional four encoders to encode up to four code cells' AST graphs. In addition, we have a high-level GRU encoder layer for all the four AST graphs to generate one high-level output. %This output is an important part of our HAConvGNN module.
More specifically, the encoder for the tokenized code sequence is a GRU with an output length of 256. An AST graph encoder is a collection of Convolutional Graph Neural Networks layers followed by a GRU layer of output length 256. We use four AST graph encoders for up to four code cells. 
Following \newcite{LeClair2019GNN},
the number of hops in our GNN layers is set to 2. %followed by Leclair et al ~\cite{LeClair2019GNN} as they demonstrated that 2 hops have the best performance.

% \subsection{Hierarchical Attention Mechanism}

% We use dot product to calculate attentions between the code sequence and the predicted documentation tokens in the decoder. In this way, we can find which tokens from the code sequence are more indicative when predicting the next token in the decoder.
%have a strong relationship with the predicted next token in our decoder. %Dot product attention computes the scores by the dot product between code sequence and decoder, which is then divided by $\sqrt{d}$ to minimize the irrelevant effect of dimension $d$ on the score. 

\subsection{HAConvGNN}
\label{sec:csam}
The key design of our HAConvGNN model is the hierarchical attention. When handling AST graphs input, instead of blending these 4 code cells as a whole sequence, we propose to use a hierarchical attention mechanism (low-level attention and high-level attention in HAConvGNN in Figure~\ref{fig:model}) on these AST graphs to better preserve the graph structure. 

Firstly, the four code cells' AST graph can be represented as $G = \{G_1, G_2, G_3, G_4\}$. We denote the decoder output (i.e., the predicted documentation tokens up till $t-1$) as $D \in \mathbb{R}^{n \times d}$ where $d$ is the dimension. We further denote each code cell's AST graph as $G_i \in \mathbb{R}^{m \times d}$ where $m$ is the number of nodes. After using a high-level encoder to encode the AST graph input, we execute a graph-level attention to get high-level attention score:
\begin{equation}
\alpha(G_i, D) = D G_i^\top / \sqrt{d}
\end{equation}
Then we apply softmax on $\alpha$, given by:
\begin{equation}
b^{i} = \frac{exp(\alpha(G_i, D))}{\sum_{j}^{ }exp(\alpha(G_j, D)))}
\end{equation}
In this way, we get the results denoted as $\alpha = \{\alpha_1, \alpha_2, \alpha_3, \alpha_4\}$. This is our high-level attention weights indicating the relations between each code cell and the already predicted documentation sequence D.

Secondly, we apply an attention mechanism on each code cell to find the relations between nodes in a code cell's AST and the predicted documentation sequence D. For each code cell's AST tree $G = \{G_1, G_2, G_3, G_4\}$, we apply the same operation as in EQ.1 and EQ.2. As a result, for each code cell $G_i$, we are able to get a new low-level attention weight $\beta_i$. For all code cells, we can denote these attention scores as $\beta = \{\beta_1, \beta_2, ..., \beta_m\}$.

Eventually, we fuse these attention weights ($\alpha$ and $\beta$) with code cells:
\begin{equation}
O = \sum_{i = 1}^{4}\alpha_i\sum_{j = 1}^{m}\beta _{i, j}G_{i,j}
\end{equation}

% We use dot product to calculate attentions between the code sequence and the predicted documentation tokens in the decoder. In this way, we can find which tokens from the code sequence are more indicative when predicting the next token in the decoder.

Now we get the AST matrices from HAConvGNN. It is then concatenated with code matrices   %from Section \ref{sec:csam}
%from the code sequence attention mechanism (Section \ref{sec:csam}) and decoder 
into a single context matrix.
Note that code matrices are based on the code sequence input with a separate uniform attention (see the left ``Code Sequence'' in Figure \ref{fig:model}).
Next, we apply a linear projection to project the merged context matrix into a 256 dimension space. This is an effective way to avoid overfitting during the training process. Finally, we flatten the new context matrix and apply another linear layer to project it into an output. The output layer size is the vocabulary size. By applying the Argmax function to the output layer, we can obtain the predicted next token (i.e., documentation token at time step $t$) in the output sequence.

\section{Experimental Setup}

\begin{table*}[t]
\centering
\renewcommand\arraystretch{1}{
\setlength{\tabcolsep}{1mm}{
\begin{tabular}{|l|c|c|c|c|c|c|c|c|c|}
\hline
\multirow{2}{*}{\texttt{\textbf{Models}}}      & \multicolumn{3}{c|}{\textbf{ROUGE-1}}                     & \multicolumn{3}{c|}{\textbf{ROUGE-2}}                  & \multicolumn{3}{c|}{\textbf{ROUGE-L}}                                     \\ \cline{2-10} 
                             & P              & R              & F1             & P             & R             & F1            & P              & R              & F1            \\ \hline
 \multicolumn{10}{|c|}{\textbf{\texttt{Baselines}}}\\ \hline
code2seq                     & 11.45          & 8.46           & 8.23           & 1.67          & 1.11          & 1.11          & 13.13          & 10.28          & 10.24          \\ \hline
graph2seq                    & 13.21          & 9.87          & 9.51          & 2.86          & 1.99          & 2.03          & 14.46          & 11.40          & 11.18         \\ \hline
 \multicolumn{10}{|c|}{\textbf{\texttt{Our Model \& Ablation Study}}}\\ \hline
HAConvGNN (Our Model)             & \textbf{22.87} & \textbf{16.92} & \textbf{16.58} & \textbf{6.72} & \textbf{4.86} & \textbf{4.97} & \textbf{24.03} & \textbf{18.60} & \textbf{18.54}  \\ \hline
\begin{tabular}[c]{@{}l@{}}HAConvGNN \\ \hspace{3mm}with low-level attention\\ \hspace{3mm}without high-level attention\\ \hspace{3mm}with uniform attention\end{tabular}     & 20.66          & 15.65           & 14.91           & 4.74          & 3.92          & 3.80          & 21.84          & 17.27           & 16.81 \\ \hline
\begin{tabular}[c]{@{}l@{}}HAConvGNN \\\hspace{3mm}with low-level attention\\  \hspace{3mm}without high-level attention\\ \hspace{3mm}without uniform attention\end{tabular}   &   19.57 & 14.59 & 14.23    &    4.87    & 3.56  & 3.63  & 20.83 & 16.24 & 16.12\\ \hline
\begin{tabular}[c]{@{}l@{}} HAConvGNN \\\hspace{3mm}without low-level attention\\ \hspace{3mm}without high-level attention \\  \hspace{3mm}with uniform attention\end{tabular}           & 11.39          & 7.73           & 7.82           & 1.58          & 1.06          & 1.08          & 13.13          & 9.47           & 9.82    \\ \hline

% w/o Attention                & 21.70          & 10.53          & 12.12          & 3.26          & 1.63          & 1.88          & 22.90          & 12.41          & 14.28          & 20.66          & 7.96           & 9.53           \\ \hline
\end{tabular}}}
\caption{ROUGE scores for the baselines, our model, and the ablation models. Results show that our model has higher scores for all three metrics, demonstrating a robust advantage over the code2seq and graph2seq models.}
\label{table:result}
\end{table*}

\subsection{Implementation Details}
We split our dataset into training, development, and test datasets at a 8:1:1 ratio. %train:dev:holdout with a 8:1:1 ratio. 
We use the Adam optimizer~\cite{Adam} with a batch size of 20. The learning rate is 0.001 and the code sequence embedding size is 100. In the encoder, we use GRU~\cite{Gru} with the hidden size of 256. The hop size of our GNN is 2. The dropout rate of our attention layer is 0.5.

% \subsection{Baseline Model}
\subsection{Baselines}
We compare our model against two baseline models which are from recent papers on the single code snippet summarization task. %Code2seq and Graph2seq are built in Keras framework. These baselines are built based on the code\footnote{\url{https://github.com/Attn-to-FC/Attn-to-FC}}.
%The reasons why we choose these models are: 1{)}. these models are from recent SOTA papers on the code summarization task; 2{)}. these models use two different approaches including using the path through the AST graph + code, and using GNN to encode the AST graph + code. 

% \textbf{Attendgru}: This is an attentional encoder-decoder model. We implemented this model followed by LeClair et al ~\cite{Leclairdata}. This is a representative attentional encoder-decoder method.

\paragraph{code2seq.} \newcite{alon2018codeseq} proposed a \emph{code2seq} model to generate a summary for a C{\#} function. 
The model creates  a vector representation for each AST path separately through an encoder. During decoding, the model uses attention to select the relevant paths. We re-implement this model and apply it on our dataset.

%They focused on C{\#} code while we focus on Python. 
%We reimplement this model in order to train on our dataset structure.

\paragraph{graph2seq.} \newcite{xu2019graphseq} proposed a graph-to-sequence learning framework that maps an input graph to a sequence of vectors and uses an attention-based LSTM method to decode the target sequence from these vectors. The authors tested the model on natural language question generation from the SQL query task. We re-implement this model using all recommended parameters from the original paper. 

%GNN based model to learn node embeddings and then reassemble them into the graph embeddings. This method proved that using GNN for AST as a separate input to the model can improve the documentation generation performance. Their work used SQL queries. To train on our dataset, We re-implement their model followed by~\cite{Haque} with all recommended parameters provided by ~\cite{xu2019graphseq}. 

%\section{Results and Discussion}
%\subsection{Automated Model Evaluation}

\subsection{Experimental Details}

The training time of code2seq model is around 2.5 hours per epoch; the training time of graph2seq is around 2.75 hours per epoch; the training time of T5-small is around 3.25 hours per epoch; the training time of our HAConvGNN model is around 2.65 hours per epoch.

The training environment of code2seq, graph2seq, and HAConvGNN is three GPUs using Parallelism. The training environment of T5-small is two GPUs.

code2seq and graph2seq are implemented in Keras framework\footnote{\url{https://github.com/Attn-to-FC/Attn-to-FC}}. T5-small model is implemented based on Huggingface repo \footnote{\url{https://github.com/huggingface/transformers}}. 

\section{Automated Evaluation}
%\textbf{Evaluation metrices}: 
We use ROUGE scores~\cite{rouge} to evaluate our model's performance with regard to the ground-truth documentation content. We report   ROUGE-1, ROUGE-2, and ROUGE-LCS (longest common sub-sequence). 
%ROUGE can be considered as a recall score to denote how much of the reference text appears in the predicted text. 
%In this paper, we report ROUGE-1, ROUGE-2, and ROUGE-LCS (longest common sub-sequence) to provide comprehensive evaluation metrics. 
%\textbf{Results}:
%The \texttt{CDG} task in a notebook is much more difficult than the previous code summary tasks. The previous works only focused on one code block and use the simple inline comment as the target. 
% Due to this difficulty, all the scores not very high. The generated average comment length is around 7, it can still prove that our model-generated summary is reasonable and meaningful. 
As shown in Table \ref{table:result}, our HAConvGNN model outperforms the other two baselines in all ROUGE metrics. 
%Therefore, it demonstrates 
%that our hierarchical model establishes a new state-of-the-art to solve code documentation generation tasks in the notebook setting.
%This highlights the importance of having the HAConvGNN structure to incorporate multiple code blocks characteristic in this context. 
%We also provide the attention visualization to help users better understand our model.

\begin{figure*}[h]
    \includegraphics[width=16cm,height=2.4cm]{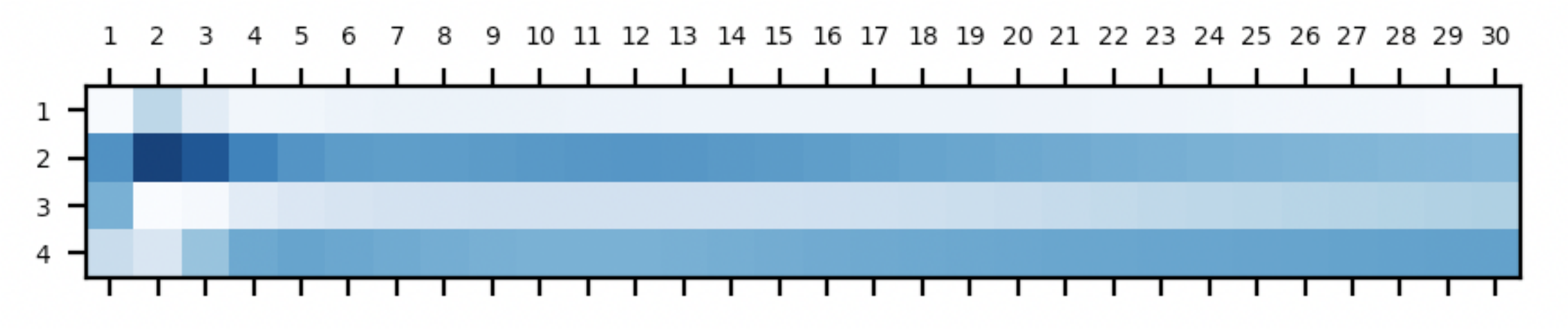}
    \caption{Attention visualization for the data point illustrated in Table~\ref{tab:example}. Each row represents a code cell, and each column is a code token. In this example, it shows the second and third token in the second code cell (`` nn\_model'', ``X'') contribute the most to the predicted documentation in Table~\ref{tab:example}.}
    \label{fig:attention}
\end{figure*}

\paragraph{Ablation study.} 
In order to better understand the impact of the attention components in our model, 
we also perform an ablation study (Table~\ref{table:result}). Our ablation study evaluates how low-level attention, high-level attention, and AST uniform attention contribute to the model. More concretely, we generate ablation models as the following: \\

\indent{(1)} without high-level attention in the hierarchical attention: we remove {high level attention} component in Figure \ref{fig:model} in our HAConvGNN structure. That means we do not compute attention weights for separated code cells.\\
\indent{(2)} without AST uniform attention: we do not apply uniform attention mechanism (i.e., the attention component above \emph{HAConvGNN} in Figure \ref{fig:model} for our HAConvGNN output with the decoder.
\indent{(3)} without low-level or high-level attentions: we remove separated low-level attention components in Figure \ref{fig:model}) in our HAConvGNN structure. Note that when we remove these separated attentions, we also remove the high-level attention (thus the entire hierarchical attention structure). We treat multiple code cells as a standalone code snippet in this situation and process graph data with the original GNN layer (see the last row in Table \ref{table:result}).\\

\begin{figure*}[htbp!]
\centering
\begin{minipage}[t]{0.32\textwidth}
\centering
\includegraphics[width=5cm]{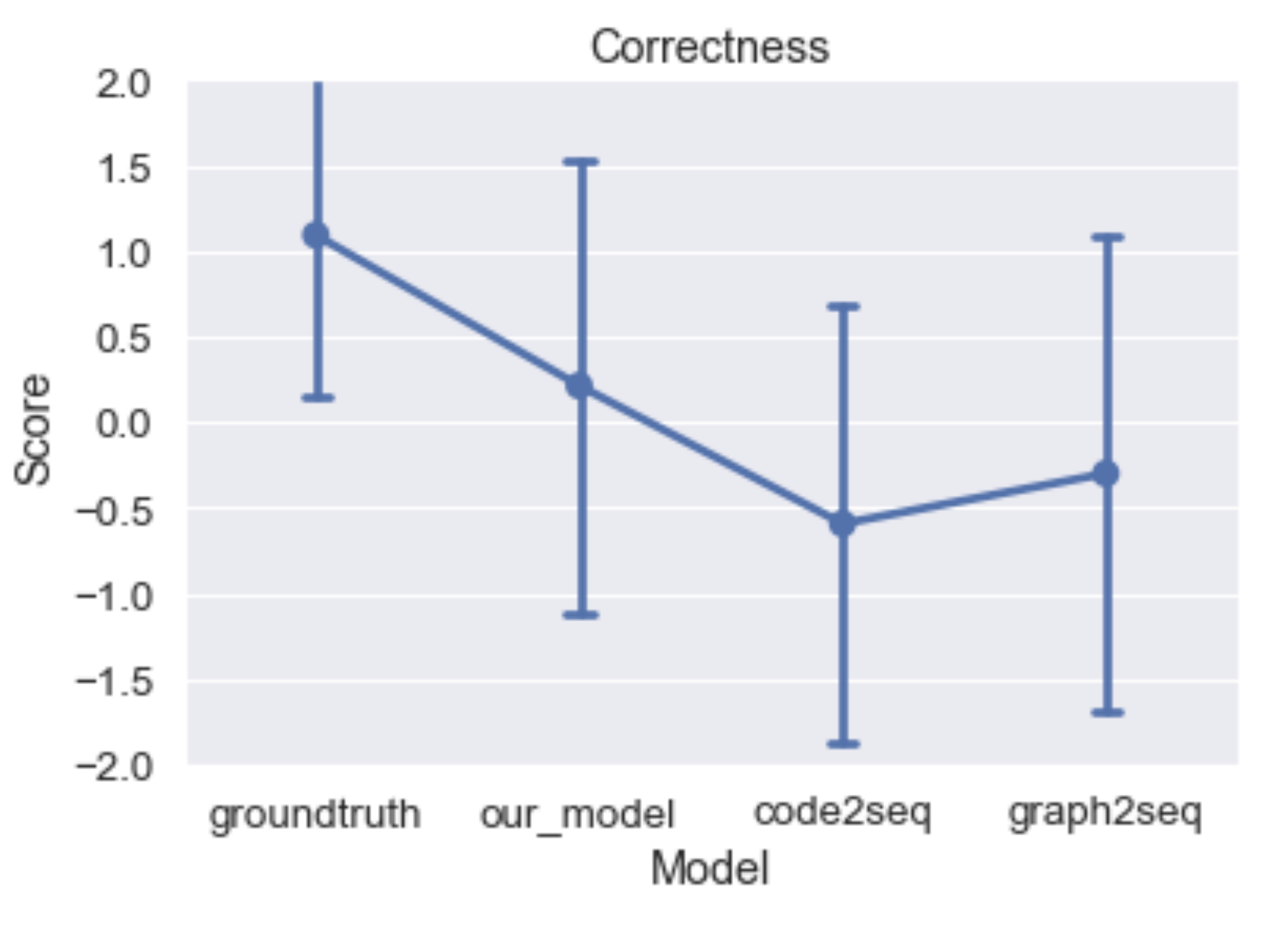}
\end{minipage}
\begin{minipage}[t]{0.32\textwidth}
\centering
\includegraphics[width=5cm]{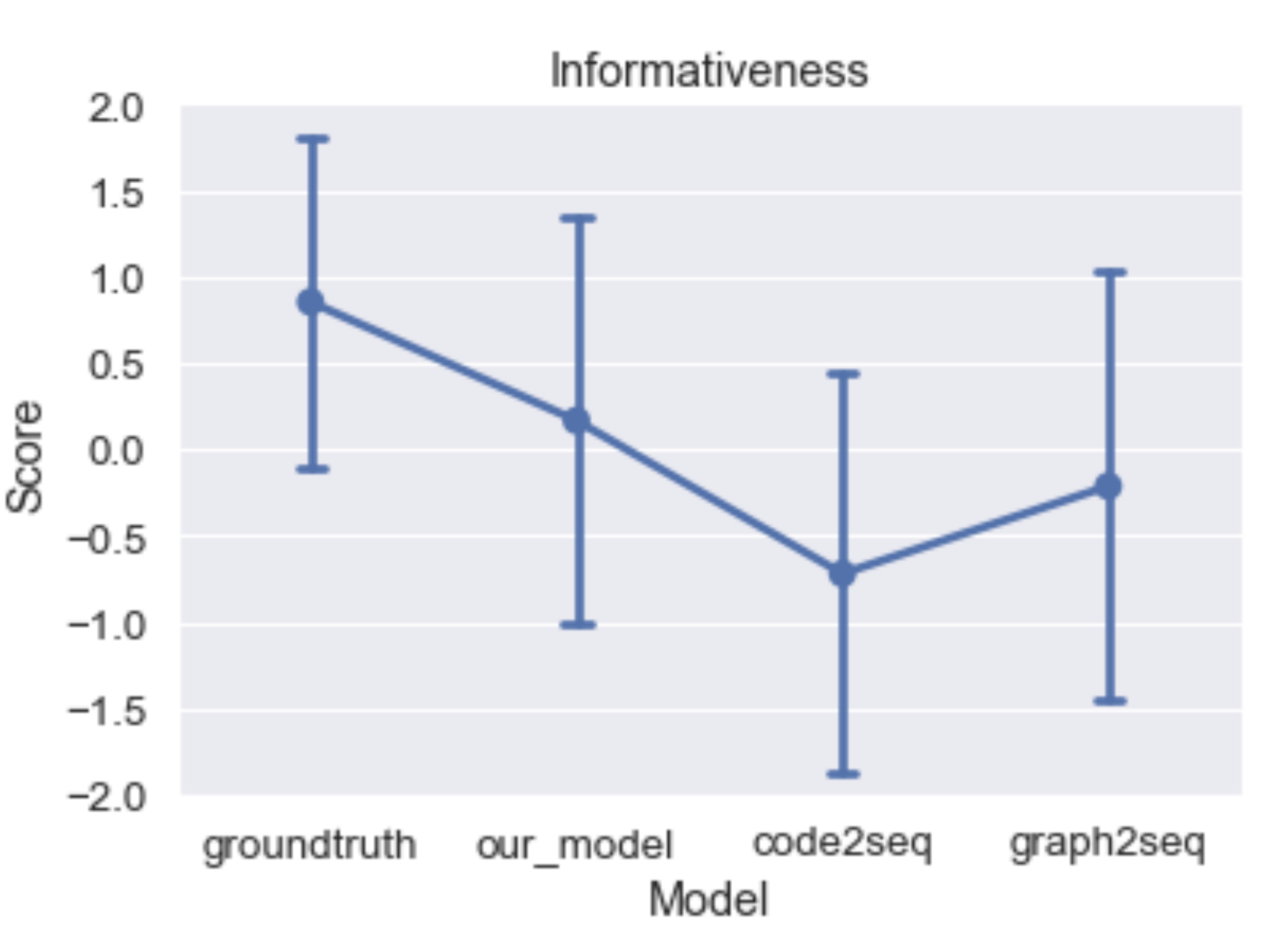}
\end{minipage}
\begin{minipage}[t]{0.32\textwidth}
\centering
\includegraphics[width=5cm]{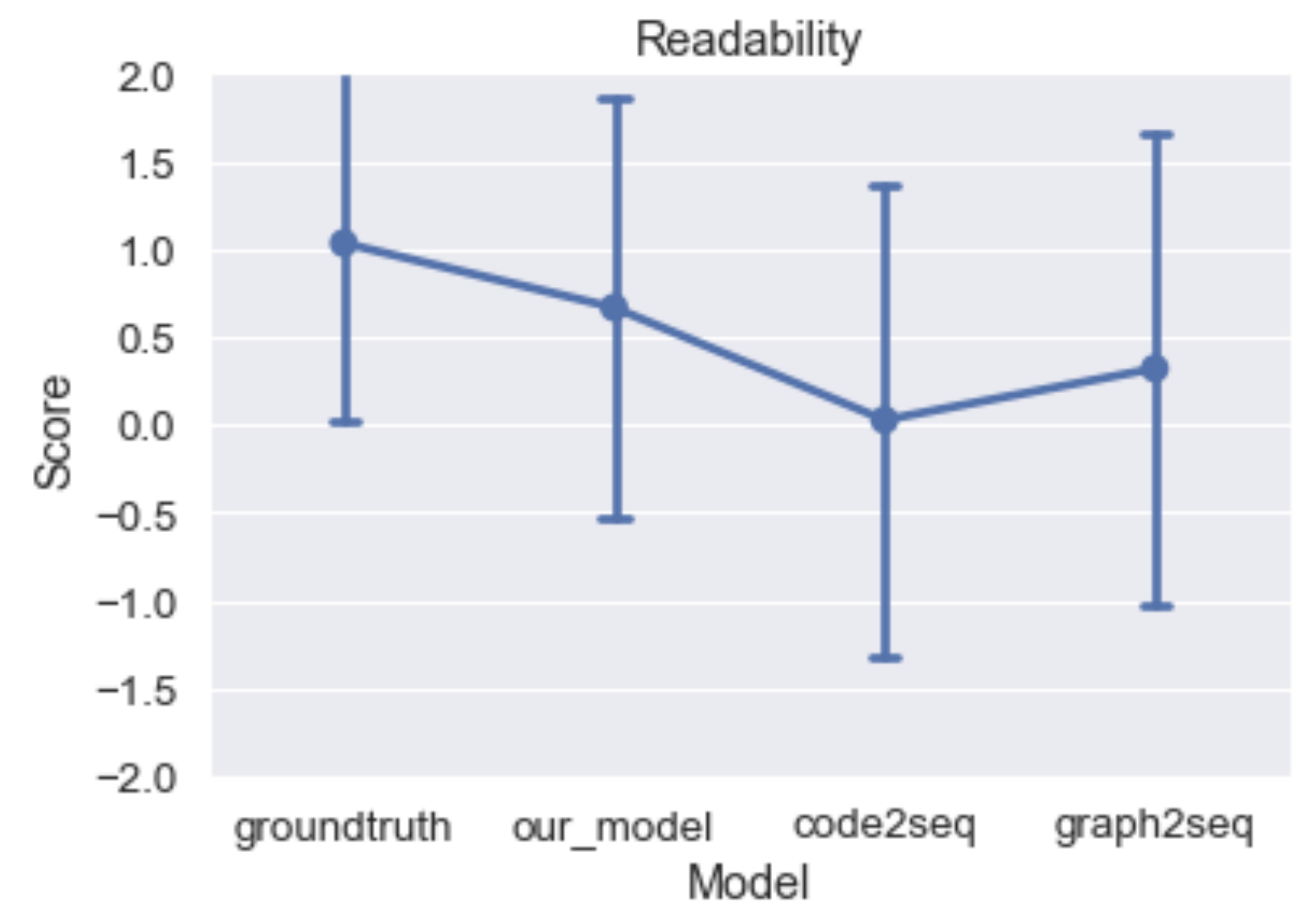}
\end{minipage}
\caption{ Average rated scores given by human evaluators to each method across three dimensions.}
\end{figure*}
\begin{table*}[ht!]
\centering
\label{tab:human-evaluation}
\small
\begin{tabular}{c c c c}
\toprule
\textbf{Model}&\textbf{Correctness}&\textbf{Informativeness}&\textbf{Readability} \\
\midrule
Groundtruth &$\overline{x}$ = 1.09, $\sigma$=0.95&$\overline{x}$ = 0.85, $\sigma$=0.97&$\overline{x}$ = 1.03, $\sigma$=1.01\\ 
Our model &$\overline{x}$ = 0.21, $\sigma$=1.33&$\overline{x}$ = 0.17, $\sigma$=1.18&$\overline{x}$ = 0.67, $\sigma$=1.20\\
Code2seq &$\overline{x}$ = -0.59, $\sigma$=1.29&$\overline{x}$ = -0.72, $\sigma$=1.17&$\overline{x}$ = 0.03, $\sigma$=1.35\\
Graph2seq &$\overline{x}$ = -0.30, $\sigma$=1.40&$\overline{x}$ = -0.21, $\sigma$=1.25&$\overline{x}$ = 0.32, $\sigma$=1.35\\
\bottomrule
\end{tabular}
\caption{Human Evaluation Result}
\label{tab:humanEval}
\end{table*}

In general, we found that the hierarchical structure in our HAConvGNN is proven to enhance our final performance. It is worth noting that the separated attention mechanism is 
%proved very useful 
essential 
in our model. Remember that we use the attention mechanism for our four code cells separately. Treating them as a single big code snippet leads to a considerable performance drop (see the last row in Table \ref{table:result}). This demonstrates that the hierarchical structure in our model can better handle the code documentation generation task for multiple code cells.

%A separated attention mechanism means that we use the attention mechanism for our four code blocks separately. %Therefore, for multiple code blocks documentation generation tasks, it is better to use the separate attention mechanism for this kind of task. According to the experiment, we can find that the hierarchical structure is very suitable and important for our task. 

%\subsection{Attention Visualization}
\paragraph{Attention Visualization.} 
Our high-level attention mechanism can indicate the most relevant code cell when generating the documentation for several code cells.  Figure~\ref{fig:attention} illustrates the attention heatmap for the code example in Table \ref{tab:example}. Note that each row represents a code cell, and each column corresponds to a code token. 
%Figure~\ref{fig:attention} shows that the second row, which is the second code block, has the darkest color thus it has the greatest influence weight on the output sequence. 
It seems that the modes pays more attention to the second code cell (especially the first few tokens)
when generating the documentation ``\emph{Implementing Neural Network}''.
% We provide a visualization (Figure~\ref{fig:attention}) to illustrate the high-level attention mechanism (correspending to the code example in Fig.~\ref{fig:code}). From Fig.~\ref{fig:code}, we speculate the second code block is the most important code blocks for the documentation ``\textit{Implementing Neural Network}''.
% Specifically, Figure~\ref{fig:attention} shows the attention weights of these code blocks which denote the relationship between these four code blocks and decoder.

% \vspace{-0.5cm}\\

\begin{figure*}[t]
    \centering
    \includegraphics[scale=0.625]{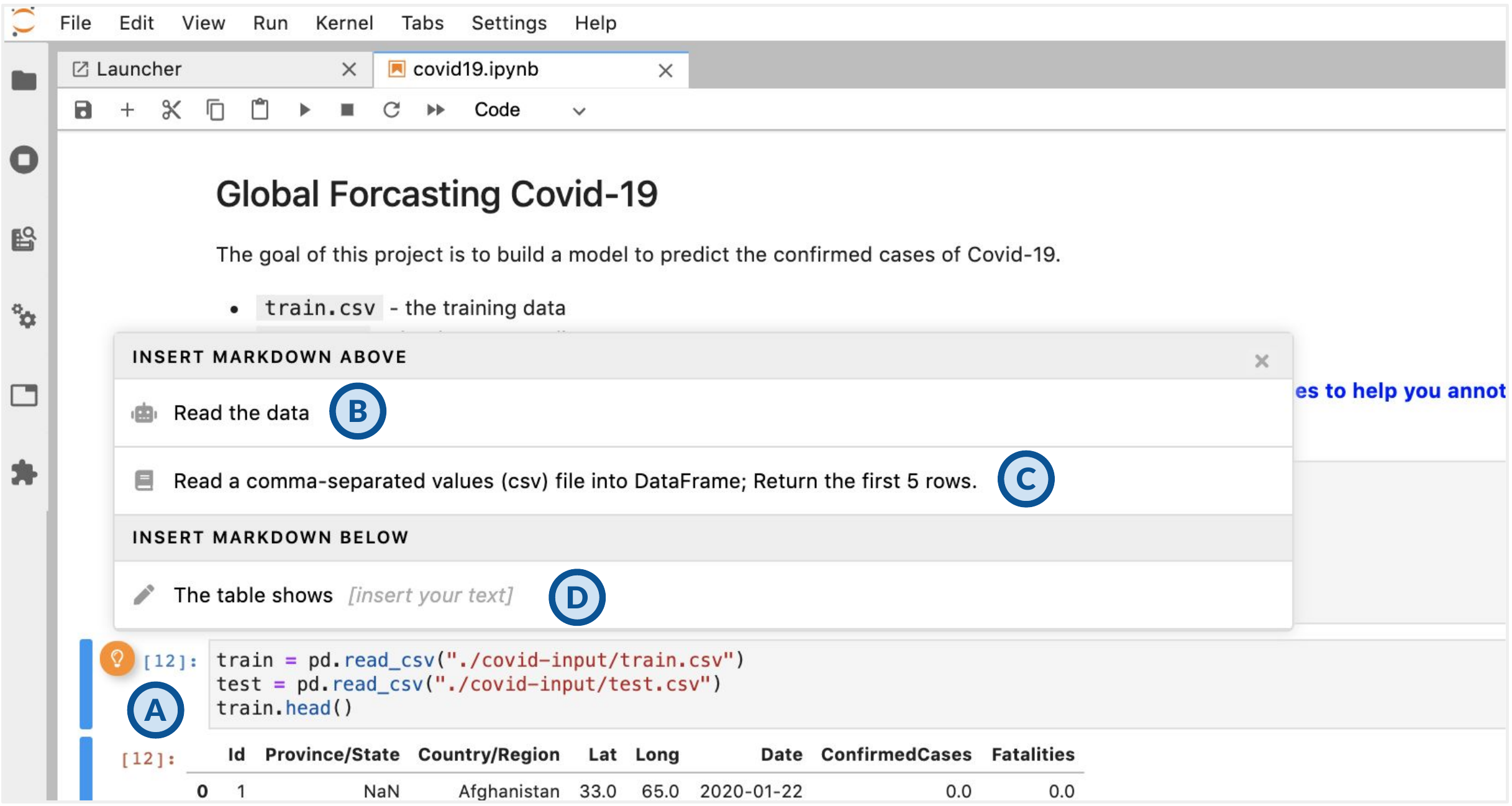}
    \caption{We implement a downstream application as a Jupyter Notebook plugin (A) to assist users documentation writing, incorporating the HAConvGNN-predicted results (B) next to an IR-based approach (C), and a user-prompt approach (D).}
    \label{fig:interface}
\end{figure*}

\section{Human Evaluation}
We also conduct a human evaluation to further evaluate our model against the two baselines and the ground truth. 
% feedback on the decided to conduct a human evaluation to ask some data engineers to evaluate documentation generated by HAConvGNN(our model) and other two baselines(code2seq and graph2seq).

\paragraph{Participants.} Our human evaluation task involves reading code snippets and rating the generated documentation of the codes. 
We recruited participants with data science and machine learning backgrounds ($N = 15$).

\paragraph{Task.} We randomly selected 30 pairs of documentation and code(s) from our dataset. Note that each pair has only one summary, but may have multiple code snippets. 
Each participant is randomly assigned 10 trials, and the order of these 10 trials is also randomized. 
Each pair is evaluated by 5 individuals. 
In each trial, a participant reads 4 candidate documentation for the same code snippet(s): three generated by the three models, and the other one is the groundtruth. 
Participants do not know which documentation text is from which model.
The participant is asked to rate the 4 documentation texts along three dimensions using a five-point Likert-scale from -2 to 2.
% \begin{itemize}
% \item Correctness:The generated summarization matches with the code content.
% \item Informativeness: The generated summarization covers more information units.
% \item Readability: The generated summarization is in readable English grammar, words etc.
% \end{itemize}
\begin{itemize}
\vspace{-10pt}
\item  \textit{Correctness}: The generated documentation matches with the code content. 
\vspace{-10pt}
\item \textit{Informativeness}: The generated documentation covers more information units.  
\vspace{-10pt}
\item \textit{Readability}: The generated documentation is in readable English grammar and words.
\end{itemize}

\paragraph{Evaluation Results.} We conducted \textit{pairwise t-tests} to compare each model's performance. 
The result (Table \ref{tab:humanEval}) shows that for the \textit{Correctness} dimension, our model (avg=0.21) is significantly better than the other two baselines (avg=-0.59 for code2seq, avg=-0.30 for graph2seq, both p<.01). Our model is also the only model that has a positive rating.
% , it suggests that the external human raters believe documentation we generate is concise and coherent.
For the \textit{Informativeness} dimension, groundtruth also has the best rating. Our model (avg=0.17) comes in second and outperforms code2seq (avg=-0.72, p<.01) and graph2seq (avg=-0.21, p<.01). 
% he rating is also over 0.

For the \textit{Readability} dimension, in which we consider whether generated documentation is a valid English sentence or not, groundtruth outperforms all ML models again, but our model (avg=0.67) also significantly outperforms baseline models code2seq (avg=0.03 p<.01) and graph2seq (avg=0.32 p<.01). Our model can generate more readable documentation than baselines.

All the results suggest that our model has above-zero ratings, which suggests it reaches an acceptable user satisfaction along all three dimensions.

\section{Comparison With Transformers}
We also carried out an additional experiment to compare our model with T5 \cite{raffel2020exploring}, which is a state-of-the-art transformer encoder-decoder model. In order to fairly compare our model against T5, we do not use any pre-trained embeddings for the T5 model.
Also, T5 input has limitation for the input token length thus we did not feed AST hierarchy into it.
More specifically, we initialize a T5-small model\footnote{In a pilot study, training a T5-base model (with random initialization) 
on our dataset leads to worse results.} with random weights and train this model using our training data. Our code adapts the transformer models from HuggingFace \cite{wolf-etal-2020-transformers}. %T5-small model is built based on the code \footnote{\url{https://github.com/huggingface/transformers}}. 
We use the dev dataset to choose the hyperparameters and evaluate the trained model on our test dataset. The ROUGE F1 scores for the trained T5-small model are as follows: ROUGE-1 = 17.55, ROUGE-2 = 4.57, ROUGE-L = 19.53. 

We found that the trained T5-small model achieves slightly better results than our model in ROUGE-1 and ROUGE-L. In practice, we found that the T5-small model relies on a much more hyperparameters and tends to generate less informative content compared to other models (see the documentation generated from different models in Table \ref{tab:example} for an example). 

But in our dataset, as reported in Table~\ref{table:dataset}, the max AST token sequence is 1,732, which is too long as T5 input (512) or BART input (1,024). That is why T5 in Sec 8 can only take the raw code sequence as input, instead of the AST hierarchy. 
It is known that programming code has a tree-based hierarchy and leveraging such AST hierarchy can enhance the baseline model (e.g.,~\cite{alon2018codeseq}).
Our contribution is that we provide a hierarchical attention architecture that is well suited for the programming code nature and can generalize to a much longer length of code inputs. 
Imagine in a scenario where we can feed a whole code repo as training input by treating each code file as a lower layer, and connecting them through function/variable referencing – our architecture can also handle that. 
In general, we think our model is orthogonal to the standard transformer models. 
One interesting future work is to integrate our hierarchical attention mechanism into the transformer-based structure instead of a GRU-based structure.

%We train a t5-small model for 15 epochs and select the model with the highest evaluated ROUGE score. As shown in Table \ref{table:result}, our HAConvGNN model outperforms the other two baselines in all ROUGE metrics.

\section{Downstream User Application}
To demonstrate the application of the HAConvGNN model, we designed a Jupyter Notebook plugin to assist document writing in data science programming (as shown in Figure \ref{fig:interface}).

The plugin is triggered when detecting users focusing on a code cell (Figure \ref{fig:interface}.A).
The plugin then reads the contents from the focused cell and its adjacent cells, and sends the content to the backend.
The backend server first generates a code summarization using the HAConvGNN model (Figure \ref{fig:interface}.B).
In addition, we implemented two other approaches to generate documentation that was intended for explaining a design decision or explaining a technical concept for educational purposes.
We retrieved the relevant documentation from the API webpage for educational purposes (Figure \ref{fig:interface}.C) and we used prompts to nudge users to explain an output (Figure \ref{fig:interface}.D).
If the user likes one of these three candidates, they can simply click on one of them, and the selected documentation candidate will be inserted into above the code cell (if it describes what and why for the code), or below it (if it interprets the result of the code).

Our plugin went through several rounds of pilot testing and iterative design. 
Participants found it reminds them to document code they would have ignored, reduce the time for developing documentation while they were actively exploring the data science task.
The implementation details and a formal evaluation of understanding the benefits of the human-AI collaborative effort for automatic documentation generation are reported separately in~\cite{wang2021themisto}.

\section{Conclusion and Future Work}

This work targets a new application that aims to automatically generate code documentation (\texttt{CDG}) for a %data science 
computational 
notebook.
This project is part of our longterm research initiative of designing AI to automated the various tasks in an AI project's lifecycle~\cite{wang2021much}.
The notebookCDG context imposes unique challenges to the current code documentation generation approaches which only consider a single code snippet.
% Due to this unique context, the existing code to summary techniques all have their limitations.
We construct a dataset from Kaggle challenge notebooks, and present a novel HAConvGNN model to encode the multiple adjacent code cells as a hierarchical AST graph to enhance a sequence model architecture. 
Both automated evaluation and human evaluation show that our model outperforms the baseline models.
%and reaches a satisfactory level. 
We also incorporate our algorithm into a Jupyter Notebook plugin to assist document writing. 

In the future, we plan to conduct more human evaluation to understand the effectiveness of our model in a real-world application scenario. 

%\newpage

\section{Ethical Concern}
Our task is an instance of natural language generation task, thus it may have potential risk and ethical issues similar to any other NLG tasks, such as the generated content may have offensive language. However, we believe our task and our approach has minimum risk of such ethical issues, due to two reasons: firstly, the language used in the context of machine learning code documentation is more strict to technical terms, offensive language is less likely to appear in the dictionary thus in our model; secondly, the dataset construction method is to use highly-voted notebooks from a publicly available Kaggle community, there is unlikely to have offensive languages in these highly-voted notebooks.
%Our next step is to further improve the model performance, so that we can incorporate such a \texttt{CDG} algorithm as an application to support data scientist documentation practice.
% In the future, we will continue this line of work and build a notebook plugin to provide the automated code documentation functionality to users (a preliminary UI design is shown in Appendix.~\ref{sec:appendix-UI}).  
% The proposed HAConvGNN model can be generalized to other software engineering contexts beyond the DS notebook task, if the input data has a hierarchical structure.
%\section*{Acknowledgements}

%Anonymized for review.
\newpage
\bibliography{anthology,custom}
\bibliographystyle{acl_natbib}

\clearpage
\appendix
\section{Appendix: Code snippets-documentation Pair Examples }
\label{sec:appendix}
\begin{table}[ht!]
\centering

\small
\begin{tabular}{|p{7.5cm}|}
\hline
\textbf{Documentation}
\\ \hline
\begin{tabular}{l l}
{\color{red} \textbf{Ground Truth}} & Feature scaling\\
{\color{blue} \textbf{Our Model}} & Feature scaling \\
{\color{magenta} \textbf{Code2seq}} & We can have the model \\
{\color{olive} \textbf{Graph2seq}} & The next step is a lot of the training set \\
{\color{cyan} \textbf{T5-small}} & Scaling \\
\end{tabular} \\
\hline
\textbf{Code Cells}
\\ \hline
\vspace{-12pt}
\begin{minted}[%
 breaklines,
 fontsize=\scriptsize
 ]{python}
from sklearn.preprocessing import StandardScaler
scaler = StandardScaler().fit(train_df)
train_scale = pd.DataFrame(scaler.transform(train_df))
\end{minted}
\\ \hline
\end{tabular}
\caption{Example: Feature Scaling}
\end{table}

\begin{table}[ht!]
\centering

\small
\begin{tabular}{|p{7.5cm}|}
\hline
\textbf{Documentation}
\\ \hline
\begin{tabular}{l l}
{\color{red} \textbf{Ground Truth}} & handle missing values in X test \\
{\color{blue} \textbf{Our Model}} & we can deal with missing values \\
{\color{magenta} \textbf{Code2seq}} & We can have the categorical data \\
{\color{olive} \textbf{Graph2seq}} & We can also make any numeric variable \\
& in the model \\
{\color{cyan} \textbf{T5-small}} & Filling the missing values in the test set \\
\end{tabular} \\
\hline
\textbf{Code Cells}
\\ \hline
\vspace{-12pt}
\begin{minted}[%
 breaklines,
 fontsize=\scriptsize
 ]{python}
cols_with_missing_val = [col for col in X_test.columns if X_test[col].isnull().any()]
print(cols_with_missing_val)
\end{minted}
\\ \hline
\vspace{-12pt}
\begin{minted}[%
 breaklines,
 fontsize=\scriptsize
 ]{python}
from sklearn.impute import SimpleImputer
my_imputer = SimpleImputer(strategy='most_frequent')
my_imputer.fit(X_train)
imputed_X_test = pd.DataFrame(my_imputer.transform(X_test))
imputed_X_test.columns = X_test.columns
\end{minted} 
\\ \hline
\end{tabular}
\caption{Example: Handle Missing Values}
\end{table}

\begin{table}[ht!]
\centering

\small
\begin{tabular}{|p{7.5cm}|}
\hline
\textbf{Documentation}
\\ \hline
\begin{tabular}{l l}
{\color{red} \textbf{Ground Truth}} & Plot the model s performance \\
{\color{blue} \textbf{Our Model}} & Plot the model s performance \\
{\color{magenta} \textbf{Code2seq}} & We can have the model \\
{\color{olive} \textbf{Graph2seq}} & The next step is a lot of the training and \\
& test set \\
{\color{cyan} \textbf{T5-small}} & Plot model performance \\
\end{tabular} \\
\hline
\textbf{Code Cells}
\\ \hline
\vspace{-12pt}
\begin{minted}[%
 breaklines,
 fontsize=\scriptsize
 ]{python}
plt.plot(history_size_val_1)
plt.plot(history_size_val_2)
plt.plot(history_size_val_3)
plt.plot(history_size_val_4)
plt.plot(history_size_val_5)
plt.plot(history_size_val_6)
plt.title('Model accuracy for different Conv sizes')
plt.ylabel('Accuracy')
plt.xlabel('Epoch')
plt.ylim(0.98,1)
plt.xlim(0,n_epochs)
plt.legend(['8-16', '16-32', '32-32', '24-48', '32-64', '48-96', '64,128'], loc='upper left')
plt.savefig('convolution_size.png')
plt.show()
\end{minted} 
\\ \hline
\end{tabular}
\caption{Example: Plot Model Performance}
\end{table}

\begin{table}[ht!]
\centering
\small
\begin{tabular}{|p{7.5cm}|}
\hline
\textbf{Documentation}
\\ \hline
\begin{tabular}{l l}
{\color{red} \textbf{Ground Truth}} & Data Augmentation \\
{\color{blue} \textbf{Our Model}} & Data Builder \\
{\color{magenta} \textbf{Code2seq}} & We can have the model \\
{\color{olive} \textbf{Graph2seq}} & LSTM \\
{\color{cyan} \textbf{T5-small}} & Visualize the images \\
\end{tabular} \\
\hline
\textbf{Code Cells}
\\ \hline
\vspace{-12pt}
\begin{minted}[%
 breaklines,
 fontsize=\scriptsize
 ]{python}
import warnings
from imgaug import augmenters as iaa
warnings.filterwarnings("ignore")

augmentation = iaa.Sequential([
        iaa.OneOf([ ## rotate
            iaa.Affine(rotate=0),
            iaa.Affine(rotate=90),
            iaa.Affine(rotate=180),
            iaa.Affine(rotate=270),
        ]),

        iaa.Fliplr(0.5),
        iaa.Flipud(0.2),

        iaa.OneOf([
            iaa.Cutout(fill_mode="constant", cval=255),
            iaa.CoarseDropout((0.0, 0.05), size_percent=(0.02, 0.25)),
            ]),

        iaa.OneOf([
            iaa.Snowflakes(flake_size=(0.2, 0.4), speed=(0.01, 0.07)),
            iaa.Rain(speed=(0.3, 0.5)),
        ]),  

        iaa.OneOf([
            iaa.Multiply((0.8, 1.0)),
            iaa.contrast.LinearContrast((0.9, 1.1)),
        ]),

        iaa.OneOf([
            iaa.GaussianBlur(sigma=(0.0, 0.1)),
            iaa.Sharpen(alpha=(0.0, 0.1)),
        ])
    ],
    random_order=True
)
\end{minted}
\\ \hline
\vspace{-12pt}
\begin{minted}[%
 breaklines,
 fontsize=\scriptsize
 ]{python}
def get_ax(rows=1, cols=1, size=7):
    _, ax = plt.subplots(rows, cols, figsize=(size*cols, size*rows))
    return ax

limit = 4
ax = get_ax(rows=2, cols=limit//2)

for i in range(limit):
    image, image_meta, class_ids,\
    bbox, mask = modellib.load_image_gt(
        dataset_train, config, image_id, use_mini_mask=False, 
        augment=False, augmentation=augmentation)
    
    visualize.display_instances(image, bbox, mask, class_ids,
                                dataset_train.class_names, ax=ax[i//2, i % 2],
                                show_mask=False, show_bbox=False)
\end{minted} 
\\ \hline
\end{tabular}
\caption{Example: Data Augmentation}
\end{table}

\begin{table}[ht]
\centering
\small
\begin{tabular}{|p{7.5cm}|}
\hline
\textbf{Documentation}
\\ \hline
\begin{tabular}{l l}
{\color{red} \textbf{Ground Truth}} & Count Monthly Mean \\
{\color{blue} \textbf{Our Model}} & Monthly Count \\
{\color{magenta} \textbf{Code2seq}} & We can have a look at the training set \\
{\color{olive} \textbf{Graph2seq}} & Feature Engineering \\
{\color{cyan} \textbf{T5-small}} & Creating a new column \\
\end{tabular} \\
\hline
\textbf{Code Cells}
\\ \hline
\vspace{-12pt}
\begin{minted}[%
 breaklines,
 fontsize=\scriptsize
 ]{python}
for year in year_list:
    for month in range(num_months_per_year): 
        start_date = datetime.datetime(year, month+1, 1, 0, 0, 0)
        end_date = datetime.datetime(year, month+1, 19, 23, 0, 0)
        count_mean = train_data[start_date:end_date]['count'].mean()
        train_data.loc[start_date:end_date, 'count_mean'] = count_mean
        
        start_date = datetime.datetime(year, month+1, 20, 0, 0, 0)
        last_day_of_month = calendar.monthrange(year,month+1)[1]
        end_date = datetime.datetime(year, month+1, last_day_of_month, 23, 0, 0)
        test_data.loc[start_date:end_date, 'count_mean'] = count_mean
\end{minted}
\\ \hline
\vspace{-12pt}
\begin{minted}[%
 breaklines,
 fontsize=\scriptsize
 ]{python}
test_data.head()
\end{minted} 
\\ \hline
\end{tabular}
\caption{Example: Count Monthly Mean}
\end{table}

\end{document}